\documentclass[aps,11pt,nofootinbib,preprintnumbers,showpacs]{revtex4}

\usepackage{amssymb, amsfonts, amsmath, bm}
\usepackage{amsfonts}
\usepackage[dvipdfmx]{graphicx,color}

\begin{document}




\title{
Supersymmetric black lenses in five dimensions
}
\vspace{2cm}

\author{Shinya Tomizawa${}^{1}$\footnote{tomizawasny`at'stf.teu.ac.jp} and Masato Nozawa${}^{2}$\footnote{nozawa`at'tap.scphys.kyoto-u.ac.jp}} 
\vspace{2cm}
\affiliation{
${}^1$ Department of Liberal Arts, Tokyo University of Technology, 5-23-22, Nishikamata, Otaku, Tokyo, 144-8535, Japan, \\
${}^2$ Department of Physics, Kyoto University, Kyoto 606-8502, Japan
}




\begin{abstract} 
We present an asymptotically flat supersymmetric black lens solution with the horizon topology $L(n,1)=S^3/\mathbb{Z}_n$ in the five-dimensional minimal ungauged supergravity. We show that the black lens carries a mass, two independent angular momenta, electric charge and $(n-1)$ magnetic fluxes, among which only the $n+1$ quantities are independent. 
\end{abstract}

\pacs{04.50.+h  04.70.Bw}
\date{\today}
\maketitle




\section{Introduction}\label{sec:1}

Black hole solutions to Einstein's equations have provided an excellent arena to test a number of classical and quantum aspects of gravity. In particular, higher dimensional black holes have attracted much attention in last two decades, for instance by the microscopic derivation of Bekenstein-Hawking entropy~\cite{Strominger:1996sh} and the realistic production of black holes at accelerators in the scenario of large extra dimensions~\cite{Argyres:1998qn}.  
In spite of startling developments for higher-dimensional black holes, our understanding of higher dimensional gravity is still poor since it is much richer and captures more degree of freedom.  According to the topology theorem of a stationary black hole in five dimensions~\cite{Hollands:2007aj,Hollands:2010qy}, 
the allowed topology of the cross section of the event horizon is restricted either to a sphere $S^3$, a ring $S^1\times S^2$ or lens spaces $L(p,q)$, provided the spacetime is asymptotically flat and allows two commuting axial Killing vector fields (more generally, the cross section of the stationary horizon must be of positive Yamabe-type under the dominant energy condition 
\cite{Galloway:2005mf,Cai:2001su}).  In the first two cases, we have the corresponding exact solutions 
to vacuum Einstein's equations~\cite{Tangherlini:1963bw,Myers:1986un,Emparan:2001wn,Pomeransky:2006bd}. In contrast, a vacuum black hole solution with the lens space topology turns out difficult to come by and is still missing.\footnote{
Although the vacuum black-hole solution constructed in \cite{Lu:2008js} has a lens space topology $L(p,q)$, it is not asymptotically flat since the spatial section at infinity has a `flipped' lens topology 
$L(q,p)$. }   Exploiting the inverse scattering method, several authors tried to find an asymptotically flat solution to the five-dimensional vacuum Einstein equations,  but unfortunately all of these attempts failed~\cite{Evslin:2008gx,Chen:2008fa}. A major obstacle in the construction of a black lens is that the resultant solutions are always plagued by naked singularities.

Recently, an asymptotically flat {\it supersymmetric} black lens solution with the topology $L(2,1)=S^3/{\mathbb Z}_2$ has been constructed  by Kunduri and Lucietti~\cite{Kunduri:2014kja} in the framework of the five-dimensional minimal ungauged supergravity (see \cite{Kunduri:2016xbo}  for the extension to ${\rm U}(1)^3$ supergravity). 
The construction relies heavily on the machinery developed by Gauntlett {\it et. al}~\cite{Gauntlett:2002nw}.  In that paper, they generalized the earlier program in \cite{Tod:1983pm} into higher dimensions and demonstrated that the supersymmetric  solutions can be systematically classified by using bilinears built out of the Killing spinor.  These bilinears define a previledged $G$-structures, which tightly constrain the possible forms of the metric and the gauge fields \cite{Gauntlett:2001ur,Gauntlett:2002sc}.  It turns out that the metric of supersymmetric solutions admitting a timelike bilinear Killing field is described in an adapted coordinate system by an $\mathbb R$-bundle over the 4-dimensional hyper-K\"ahler base space and the governing equations are linear, allowing us to find a number of exact solutions of physical interest.  The program undertaken in~\cite{Gauntlett:2002nw} has replaced the standard ansatz-based approaches and provoked a drastic progress in classifications of BPS solutions with various fractions of supersymmetry in diverse supergravities~\cite{Gutowski:2004yv,Gutowski:2005id,Bellorin:2006yr,Bellorin:2007yp,Gutowski:2003rg,Gauntlett:2002fz,Gauntlett:2003wb,Caldarelli:2003pb,Bellorin:2005zc,Meessen:2010fh,Meessen:2012sr,Nozawa:2010rf,Klemm:2015mga}. 
An alternative approach referred to as a spinorial geometry also gives us an elegant and  powerful method for classifying supergravity solutions (see~\cite{Gran:2005wn,Gran:2005ct,Cacciatori:2007vn,Gutowski:2007ai,Grover:2008ih,Cacciatori:2008ek,Klemm:2009uw,Klemm:2010mc} for an incomplete list of references).  These domains of research have been motivated by the desire to figure out the string duality and the gauge/gravity correspondence.

For the supersymmetric black objects in asymptotically flat spacetimes in the five-dimensional minimal supergravity, a number of properties have been clarified by many authors. The most salient feature of a supersymmetric black hole is that the horizon must be degenerate and non-rotating. 
This is because the bilinear Killing vector becomes never spacelike, and hence the ergoregion does not exist~\cite{Gauntlett:1998fz}. Reall gave a proof that the possible topologies of the supersymmetric black holes are either $S^3$, $S^1\times S^2$, $T^3$ or quotient thereof~\cite{Reall:2002bh}. 
For the first case,  Breckenridge {\it et al.} constructed a 
black hole with a spherical topology admitting angular momenta~\cite{Breckenridge:1996is}, referred to  as a BMPV black hole. The BMPV black hole is characterized by the mass and two equal angular momenta, 
due to which the spatial symmetry is enhanced to ${\rm U}(2)\simeq {\rm SU}(2)\times {\rm U}(1)$~\cite{Gibbons:1999uv}.
As in the Kerr case, the magnitude of angular momentum is bounded by the mass, otherwise a naked time machine shows up. 
The second possibility corresponding to a black hole with a ring topology $S^1\times S^2$ has been found in \cite{Elvang:2004rt}.
The black ring enjoys only the ${\rm U}(1)\times {\rm U}(1)$ spatial symmetry and precludes a configuration with equal angular momenta, which distinguish it from the BMPV black hole. The angular momentum along the $S^1$ direction does not have an upper bound, in sharp contrast to the spherical counterpart. A more intriguing feature arises in the ${\rm U}(1)^3$ supergravity, where the three-charge black ring is specified by 7 parameters, while only 5 of them are conserved quantities~\cite{Elvang:2004ds,Bena:2004de}. This exhibits a classical infinite non-uniqueness of black objects.

Ref.~\cite{Reall:2002bh} also proved that the only asymptotically flat black-hole solution to five-dimensional minimal supergravity whose near-horizon geometry is locally isometric to that of BMPV black hole is the BMPV 
black hole, provided that the bilinear Killing field is everywhere timelike outside the Killing horizon.  
The black lens space found in~\cite{Kunduri:2014kja} does not counter to the above uniqueness theorem, 
since the bilinear Killing field happens to be null at some points outside the horizon. 
These ``critical surfaces'' or ``evanescent ergosurfaces'' \cite{Gibbons:2013tqa} are regular timelike surfaces and provide us with much richer varieties of black holes. For instance, ref.~\cite{Kunduri:2014iga} pointed out that the black hole with a spherical topology other than the BMPV solution indeed exists. Another interesting facet of a black lens in~\cite{Kunduri:2014kja} and the solution in \cite{Kunduri:2014iga} is that they have nontrivial 2-cycles outside the horizon. This kind of ``bubbling'' solutions is of crucial importance in the context of fuzzball conjectures~\cite{Bena:2007kg}, and a number of solitons and black holes/rings with bubbles have been found~\cite{Bena:2009qv,Compere:2009iy,Bobev:2009kn,Bena:2005va}. A key ingredient to bring about these bubbles is that the spatial hypersurface has nontrivial second homology class, which is not realizable in four dimensions.

In this paper, we generalize the work of Kunduri and Lucietti~\cite{Kunduri:2014kja} to the more general lens space, and  construct an asymptotically flat supersymmetric black lens solution with the horizon topology of $L(p,1)=S^3/{\mathbb Z}_p$ for $p\ge 3$ in the five-dimensional minimal ungauged supergravity.   The regular metric on the lens space 
$L(p,1)=S^3/{\mathbb Z}_p$ with unit radius can be written as 
\begin{eqnarray}
ds^2=\frac 14 \left[\left(\frac{d\psi}{p}+\cos\theta d\phi\right)^2+d\theta^2+\sin^2\theta d\phi^2\right] ,
\label{lens}
\end{eqnarray}
where $0\le \psi<4\pi$, $0\le \phi<2\pi$ and $0\le \theta\le\pi$. 
The parameter $p$ is an integer parametrizing the Chern class of the principal bundle over $S^2$. 
In particular, this reduces to a metric on a three-dimensional sphere for $p=1$ written in terms of the Euler angle coordinates. 
Our strategy is to consider the Gibbons-Hawking space as a hyper-K\"ahler base space and allow the harmonic functions to have $n$ point sources with appropriate coefficients. By imposing suitable boundary conditions, we find the configuration in which the cross section of the horizon becomes $p=n\ (n\ge 3)$ and the cross section of null infinity becomes $p=1$. Our black lens solution possesses nontrivial 2-cycles supported by magnetic fluxes outside the horizon. One of them touches the horizon and forms the disk topology, while others are away from the horizon.

\medskip
We organize the present paper as follows. 
In the next section~\ref{sec:solution}, we present the supersymmetric solutions 
describing black lenses in the five-dimensional minimal ungauged supergravity. 
The solution is stationary and bi-axially symmetric admitting ${\rm U}(1)\times {\rm U}(1)$ isometry.  
In section~\ref{sec:boundary}, we study in depth the boundary conditions under which the spacetime 
is asymptotically flat, no closed timelike curves  (CTCs) appear around the horizon, 
no (conical and curvature) singularities develop in the domain of outer communications, 
and no orbifold singularities nor Dirac-Misner strings exist on the axis.  This boundary conditions
place restrictions upon the parameters and it turns out that the physical solution is specified by $n+1$ parameters. Section~\ref{sec:analysis} analyzes some physical properties of black lenses. 
This includes the discussion about conserved quantities and CTCs.     
 In the final section~\ref{sec:discuss}, we devote ourselves to the summary and discussion on our results.




\section{Black lens solution}
\label{sec:solution}

Let us begin with a basic setup for supersymmetric solutions in the five-dimensional minimal ungauged supergravity, whose bosonic Lagrangian consists of the Einstein-Maxwell theory with  a Chern-Simons term and takes the form~\cite{Cremmer}
\begin{eqnarray}
\mathcal L=R \star 1 -2 F \wedge \star F -\frac 8{3\sqrt 3}A \wedge F \wedge F \,, 
\end{eqnarray}
where $F=d A$ is the Maxwell field. 
The gravitational solution is said to be supersymmetric if it admits a spinor 
obeying the first-order differential equations
\begin{eqnarray}
\hat \nabla_\mu \epsilon:=\left(\nabla_\mu +\frac {i}{4\sqrt 3}(\gamma_{\mu\nu\rho}-4g_{\mu\nu}\gamma_\rho )
F^{\nu\rho} \right)\epsilon=0\,.
\label{KS}
\end{eqnarray}
The supersymmetric solutions to five-dimensional minimal ungauged supergravity have been systematically classified according to the causal nature of the Killing vector $V^\mu=i \bar \epsilon \gamma^\mu \epsilon$ constructed out of the Killing spinor~\cite{Gauntlett:2002nw}. In the domain where $V$ is timelike, the local metric and the gauge field strength have a simple description in the $t$-independent form 
\begin{eqnarray}
\label{metric}
ds^2&=&-f^2(dt+\omega)^2+f^{-1}ds_{M}^2,\\
F&=&\frac{\sqrt{3}}{2}d[f(dt+\omega)]-\frac 1{\sqrt 3}G^+,
\end{eqnarray}
where $V=\partial/\partial t$ and $ds^2_M$ is the metric of a hyper-K\"{a}hler base space. 
The norm $f$ and the twist $\omega $ of a Killing vector $V$  are a scalar and a 1-form on
the base space and obey a linear system
\begin{eqnarray}
\Delta_h f^{-1} =\frac 29 (G^+)_{mn}(G^+)^{mn} \,, \qquad
G^+ =\frac 12 f(d\omega+ \star _h d \omega) \,, \qquad d G^+=0\,, 
\end{eqnarray}
where we have employed a convention that the hypercomplex structures are anti-self-dual. 
If this system is solved, the solution to the Killing spinor equation (\ref{KS}) is given by $\epsilon=f^{1/2} \eta$, 
where $ \eta$ is a covariantly constant chiral Killing spinor of the hyper-K\"ahler base space, i.e., 
the solution preserves at least half of supersymmetries. 

Among a variety of hyper-K\"ahler spaces, the Gibbons-Hawking space~\cite{Gibbons:1979zt} plays a distinguished role. 
The metric of the Gibbons-Hawking space reads
\begin{eqnarray}
ds^2_M&=&H^{-1}(d\psi+\chi)^2+Hdx^idx^i, \qquad 
d \chi = * d H \,,
\end{eqnarray}
where $\{x^i\}=(x,y,z)\ (i=1,2,3)$ are Cartesian coordinates on $\mathbb{E}^3$ and 
$\partial/\partial \psi$ is a triholomorphic Killing vector. When 
$\partial/\partial \psi$ continues a symmetry generator for the five-dimensional metric $g_{\mu \nu}$ and the gauge field $A_\mu$, 
it commutes with $V$ and the supersymmetry-preserving dimensional reduction is possible.\footnote{
When the Kaluza-Klein Killing vector fails to commute with the supersymmetric Killing vector, 
the dimensional reduction breaks supersymmetry. See~\cite{Klemm:2015qpi} for a notable example.} 
Furthermore, every bosonic element can be obtained in a closed form and reads~\cite{Gauntlett:2002nw,Gauntlett:2004wh} 
\begin{eqnarray}
f^{-1}&=&H^{-1}K^2+L,\\
\omega&=&\omega_\psi(d\psi+\chi)+\hat \omega,\\
\omega_\psi&=&H^{-2}K^3+\frac{3}{2} H^{-1}KL+M, \\
*d\hat\omega&=&HdM-MdH+\frac{3}{2}(KdL-LdK),\\
G^+&=& \frac 32 d \left[\frac KH (d \psi +\chi )+\xi\right] \,, \qquad d \xi =-* d K \,,
\end{eqnarray}
which leads to 
\begin{eqnarray}
A=\frac{\sqrt 3}{2} \left[f(d t+\omega)-\frac KH(d \psi+\chi)-\xi \right]\,. 
\end{eqnarray}
Here $H, K, L, M$ are harmonic functions on $\mathbb E^3$, which are only the necessary input to specify the supersymmetric solutions.  It should be observed that there exists a gauge freedom of redefining harmonic functions~\cite{Bena:2005ni} 
\begin{eqnarray}
K\to K+aH,\qquad L\to L-2aK-a^2H,\qquad M\to M-\frac{3}{2}aL+\frac{3}{2}a^2K+\frac{1}{2}a^3H,\label{eq:trans}
\end{eqnarray}
where $a$ is a constant. Under (\ref{eq:trans}), one can easily verify that 
($f, \omega_\psi, \chi$) remain invariant, whereas the 1-form $\xi$ undergoes a change as
$\xi\to \xi-a\chi$. 
Since this transformation merely amounts to the gauge transformation 
$A\to A+a d \psi$,  the transformation (\ref{eq:trans}) makes the bosonic sector invariant. 
We shall come back to this freedom (\ref{eq:trans}) in the following analysis. 

When the general choice of these harmonics are made, the solution fails to be asymptotically flat or 
suffers from CTCs, and the existence of the horizon is not guaranteed.  Following the paper by Kunduri and Lucietti~\cite{Kunduri:2014kja} (see also \cite{Crichigno:2016lac}), we consider the following class of harmonic functions
\begin{eqnarray}
H&=&\sum_{i=1}^n\frac{h_i}{r_i}:=\frac{n}{r_1}-\sum_{i=2}^n\frac{1}{r_i}, \label{Hdef}\\
M&=&m_0+\sum_{i=1}^n\frac{m_i}{r_i},\label{Mdef}\\
K&=&\sum_{i=1}^n\frac{k_i}{r_i},\label{Kdef}\\
L&=&l_0+\sum_{i=1}^n\frac{l_i}{r_i}.\label{Ldef}
\end{eqnarray}
Here, $r_i:=|{\bm r}-{\bm r_i}|=\sqrt{(x-x_i)^2+(y-y_i)^2+(z-z_i)^2}$, where $(x_i,y_i,z_i)$ are constants. 
This choice of harmonics reduces to that of the BMPV black hole~\cite{Breckenridge:1996is} for $n=1$, and that of the
black lens with $S^3/{\mathbb Z}_2$ for $n=2$~\cite{Kunduri:2014kja}.  

The 1-forms ($\chi, \xi, \hat \omega$) are obtained by
\begin{eqnarray}
\chi&=&\sum_{i=1}^nh_i\tilde\omega_i,\\
\xi&=-&\sum_{i=1}^nk_i\tilde\omega_i,\\
\hat \omega&=&\sum_{i,j=1(i\not=j)}^n\left(h_im_j+\frac{3}{2}k_i l_j \right)\hat \omega_{ij}-\sum_{i=1}^n\left(m_0h_i+\frac{3}{2}l_0k_i\right)\tilde\omega_i.
\end{eqnarray}
where 1-forms 
$\tilde\omega_{i}$ and $\hat \omega_{ij}$ ($i\not=j$) on ${\mathbb E}^3$ are defined by (c.f~\cite{Dunajski:2006vs}).
\begin{eqnarray}
\hat\omega_{ij}&=&-\left(\frac{({\bm r}-{\bm r}_i)\cdot ({\bm r}-{\bm r}_j)}{r_ir_j}+c_{ij}\right)\frac{\left[ ({\bm r}_i-{\bm r}_j)\times ({\bm r}-\frac{{\bm r_i}+{\bm r_j}}{2})\right]_kdx^k}{\left|({\bm r}_i-{\bm r}_j)\times({\bm r}-\frac{{\bm r_i}+{\bm r_j}}{2})\right|^2},\\
\tilde\omega_i&=&
\frac{z-z_i}{r_i}
\frac{(x-x_i)dy-(y-y_i)dx}{(x-x_i)^2+(y-y_i)^2},
\end{eqnarray}
where  $c_{ij}=-c_{ji}$ are constants.\footnote{
One can also add pure-gradient terms to $\tilde\omega_i$. However, 
the regularity of the solution requires that these additional constants must vanish in the present context. } 
These 1-forms satisfy $*d\tilde\omega_i=d(1/r_i)$ and  
$*d\hat \omega_{ij}=(1/r_i)d(1/r_j)-(1/r_j)d(1/r_i)$.
Throughout this paper, we set $x_i=y_i=0$ for all $i$ (in this case
$x\partial/\partial y-y\partial/\partial x$ is another Killing field) and assume $z_i<z_j$ for $i<j$.
In this case, 
$\hat \omega$ and $\tilde \omega_i$ are simplified in  spherical  coordinates 
($x=r\sin\theta\cos\phi, y=\sin\theta\sin\phi, z=r\cos\theta$) to 
\begin{eqnarray}
\hat \omega&=&\Biggl[\sum_{i,j=1(i\not=j)}^n\left(h_im_j+\frac{3}{2}k_i l_j \right)\frac{r^2-(z_i+z_j)r\cos\theta+z_iz_j}{z_{ji}r_ir_j} \notag \\
&& -\sum_{i=1}^n\left(m_0h_i+\frac{3}{2}l_0k_i\right)\frac{r\cos\theta-z_i}{r_i}+c\Biggl]d\phi \,, 
\\
\tilde \omega_i &=&\frac{z-z_i}{r_i}d \phi \,,  
\end{eqnarray}
with the constant
\begin{eqnarray}
c:=\sum_{i,j=1(i\not=j)}^n\left(h_im_j+\frac{3}{2}k_il_j\right)\frac{c_{ij}}{z_{ji}}\,,
\end{eqnarray}
where $z_{ji}:=z_j-z_i$. Obviously, we have a freedom to choose $z\to z+{\rm const.}$, 
which can be used to set $z_1=0$ without losing any generality. 

Under the transformation (\ref{eq:trans}),  
the coefficient of $1/r_1$ term in $M$ varies as 
$m_1\to m_1+\frac 12 (a^3+3k_1a^2-3l_1a)$.  
It therefore turns out that an appropriate choice of $a$ allows us to  set 
\begin{eqnarray}
m_1=0.  \label{eq:m1}
\end{eqnarray}
Hence, the solution contains $4n+1$ parameters 
($c, l_0, m_0, k_i, l_i, m_{i\ge 2}, z_{i\ge 2}$). We shall see below that 
the regular boundary conditions constrain some of them and the physical solution is specified only by 
$n+1$ parameters.




\section{Boundary conditions}\label{sec:boundary}

In order to obtain a supersymmetric black lens solution of physical interest, 
we impose suitable boundary conditions at (i) infinity, (ii) horizon ${\bm r}={\bm r}_1$, 
(iii) ``bubbles'' ${\bm r}={\bm r}_i\ (i=2,...,n)$ and (iv) axis $x=y=0$.  
The boundary condition at infinity (i) must be such that the spacetime is asymptotically flat. 
At the horizon (ii), the surface ${\bm r}={\bm r}_1$ should correspond to a smooth degenerate null surface whose 
spatial cross section has a topology of the lens space 
$L(n,1)=S^3/{\mathbb Z}_n$.  
At  the $(n-1)$ points ${\bm r}={\bm r}_i\ (i=2,...,n)$ (iii) where each harmonic function diverges,  
we put constraints upon parameters in such a way that these points correspond merely to the coordinate singularities like the origin of the Minkowski spacetime. 
On the axis, we demand that there appear no Dirac-Misner strings, and  
orbifold singularities at isolated points must be eliminated. 
At these boundaries,  the spacetime is required to allow neither CTCs nor 
(conical and curvature) singularities.




\subsection{Infinity}
\label{sec:bc_infinity}

Let us begin by addressing the asymptotic flatness of the solution. 
For $r \to \infty$, the metric functions ($f, \omega_\psi$) behave as 
\begin{eqnarray}
f^{-1}&\simeq &l_0+\left[\left(\sum_{i}k_i\right)^2+\sum_{i}l_i\right] r^{-1}, \qquad 
\omega_\psi\simeq m_0+\frac{3}{2}l_0\sum_{i} k_i. 
\end{eqnarray}
Since the 1-forms $\tilde\omega_{i}$ and $\hat\omega_{ij}$ are approximated by
\begin{eqnarray}
\tilde \omega_i \simeq \cos\theta d\phi,\qquad \hat\omega _{ij}\simeq \frac{d\phi}{z_{ji}}.
\end{eqnarray}
one gets
\begin{eqnarray}
\chi&=& \sum_{i} h_i \hat\omega_i\simeq \sum_{i} h_i\cos\theta d\phi=\cos\theta d\phi,\\
\omega
                    &\simeq & \left(m_0+\frac{3}{2}l_0\sum_{i}k_i\right)\left(d\psi +\cos\theta d\phi \right)-\sum_{i}\left(m_0h_i+\frac{3}{2}l_0k_i\right)\cos\theta d\phi \notag \\
                    &&+\left(\sum_{i,j(i\not =j)}\frac{h_im_j
                    +\frac{3}{2}k_il_j}{z_{ji}} +c\right)d\phi.
\end{eqnarray}
The asymptotic flatness demands the parameters to satisfy
\begin{eqnarray}
l_0&=&1,\label{eq:l0}\\
c&=&-\sum_{i,j(i\not =j)}\frac{h_im_j+\frac{3}{2}k_il_j}{z_{ji}},\label{eq:c}\\
m_0&=&-\frac{3}{2}\sum_{i}k_i.\label{eq:m0}
\end{eqnarray}
In terms of the radial coordinate $\rho=2\sqrt{r}$, 
the metric asymptotically ($\rho \to \infty$) approaches to 
\begin{eqnarray}
ds^2&\simeq& -dt^2+d\rho^2+\frac{\rho^2}{4}\left[(d\psi+\cos\theta d\phi)^2+ d\theta^2+\sin^2\theta d\phi^2 \right].
\end{eqnarray}
This is nothing but the metric of Minkowski spacetime for which the $S^3$ corresponding to the cross section of spatial infinity is written in terms of Euler angles ($\psi, \phi, \theta)$. 
The avoidance of conical singularities requires the range of angles to be 
$0\le \theta \le \pi$, $0\le \phi <2\pi $ and $0\le \psi <4\pi$ with the identification
$\phi\sim \phi+2\pi$ and $\psi\sim \psi+4\pi$.




\subsection{Horizon}

We next show that the point source ${\bm r}={\bm r}_1$ corresponds to a degenerate Killing horizon and the  topology of the spatial cross section is a lens space $L(n,1)=S^3/\mathbb{Z}_n$. 
Since we have imposed $z_1=0$, we now look at the behavior of the solution around 
the origin ${\bm r}=0$. Since four harmonic functions $H$, $K$, $L$ and $M$ are
expanded as 
\begin{eqnarray}
&&H\simeq  \frac{h_1}{r}+\sum_{i\not=1}\frac{h_i}{|z_{i1}|},\qquad  K\simeq\frac{k_1}{r}+\sum_{i\not=1}\frac{k_i}{|z_{i1}|}, \notag \\
&&L\simeq\frac{l_1}{r}+l_0+\sum_{i\not=1}\frac{l_i}{|z_{i1}|},\qquad  M\simeq m_0+\sum_{i\not=1}\frac{m_i}{|z_{i1}|},
\end{eqnarray}
the functions $f^{-1}$ and $\omega_\psi$ reduce to 
\begin{eqnarray}
f^{-1}\simeq \frac{k_1^2/n+l_1 }{r}+c_1',\qquad \omega_\psi &\simeq&\frac{k_1^3/n^2+3k_1l_1/2n}{r}+c_2'. 
\end{eqnarray}
Here we have defined the constants $c_1'$ and $c_2'$ by
\begin{eqnarray}
c_1'&:=&l_0+\sum_{i\not=1}\frac{1}{h_1^2|z_{i1}|}[2h_1k_1k_i-k_1^2h_i+h_1^2l_i],\\
c_2'&:=&m_0+\frac{3}{2h_1}k_1l_0\nonumber\\
&&+\sum_{i\not =1}\frac{1}{2h_1^3|z_{i1}|}[-(4k_1^3+3h_1k_1l_1)h_i+3h_1(2k_1^2+h_1l_1)k_i+3h_1^2k_1l_i+2h_1^3m_i].
\end{eqnarray}
The $1$-forms $\tilde\omega_i$ and $\hat\omega_{ij}$ are 
\begin{eqnarray}
&&\tilde\omega_1\simeq \cos\theta d\phi,\qquad \tilde\omega_i\simeq -\frac{z_{i1}}{|z_{i1}|} d\phi \ (i\not=1),\\
&&\hat\omega_{1j}\simeq -\frac{\cos\theta}{|z_{j1}|} d\phi \ (j\not=1),\qquad \hat\omega_{ij}\simeq \frac{z_{i1}z_{j1}}{|z_{i1}z_{j1}|z_{ji}} d\phi \ (i,j\not=1,i\not = j), 
\end{eqnarray}
yielding 
\begin{eqnarray}
\hat\omega
&=&  \biggl[\sum_{j\not=1}\left(nm_j+\frac{3}{2}k_1 l_j- \frac{3}{2}k_jl_1\right)\frac{-\cos\theta}{|z_{j1}|} + \sum_{i,j\not=1(i\not=j)}\left(-m_j+\frac{3}{2}k_i l_j \right) \frac{z_{i1}z_{j1}}{|z_{i1}z_{j1}|z_{ji}}  \nonumber\\
&&-\left(m_0n+\frac{3}{2}l_0k_1\right)\cos\theta -\sum_{i\not=1}\left(-m_0+\frac{3}{2}l_0k_i\right) \frac{-z_{i1}}{|z_{i1}|}+c \biggr]d\phi,
\end{eqnarray}
and
\begin{eqnarray}
\chi=  h_1 \hat\omega_1+\sum_{i\not=1} h_i \hat\omega_i\simeq \left(n\cos\theta+\sum_{i\not=1}\frac{z_{i1}}{|z_{i1}|} \right)d\phi.
\end{eqnarray}
In terms of new coordinates $(v,\psi')$ given by
\begin{eqnarray}
dv=dt-\left(\frac{A_0}{r^2}+\frac{A_1}{r}\right)dr,\qquad 
d\psi'=d\psi+\sum_{i\not=1}\frac{z_{i1}}{|z_{i1}|}d\phi-\frac{B_0}{r}dr.
\end{eqnarray}
we wish to require that the metric is regular around ${\bm r}=0$. 
A potential divergence appears from $g_{rr}$ and $g_{r\psi'}$, which can be 
eliminated by  
 \begin{eqnarray}
 {A_0}&=&\frac{1}{2} \sqrt{3  {k_1}^2  {l_1}^2+4 n  {l_1}^3},\\
 {A_0B_0}&=&\frac{2  {k_1}^3+3  {k_1}  {l_1} n }{2},\\
 {4A_0A_1}&=&-4m_0k_1^3+3l_1k_1^2-6nk_1l_1m_0+6nl_1^2\nonumber\\
 &+&\sum_{i\not=1}\frac{1}{|z_{i1}|}[3k_1 l_1^2k_i+3(k_1^2 l_1 + 2n l_1^2 )l_i
 -2(2 k_1^3 + 3n k_1 l_1)m_i-2l_1^3].
 \end{eqnarray}
 With this choice, it turns out that the metric is then analytic in $r $ and therefore can be extended into the $r<0$ region. It follows that the null surface $r=0$ corresponds to the Killing horizon for the supersymmetric Killing field 
$V=\partial/\partial v$. 

Taking the higher-order terms in $r$ into account, one obtains the near-horizon limit 
by  $(v,r)\to (v/\epsilon,\epsilon r)$ and $\epsilon \to 0$ \cite{Reall:2002bh}. After some algebra, 
one arrives at  
\begin{eqnarray}
ds^2_{\rm NH}&=&\frac{R_2^2}{4}\left[d\psi'+n\cos\theta d\phi -\frac{2k_1(2k_1^2+3nl_1)}{R_1^4R_2^2}rdv\right]^2+R_1^2(d\theta^2+\sin^2\theta d\phi^2)\nonumber \\
&&-\frac{4r^2}{R_1^2R_2^2}dv^2-\frac{4}{R_2}dvdr,
\label{NHmetric}
\end{eqnarray}
and 
\begin{eqnarray}
A=\frac{\sqrt{3}}{2}\left[\frac{nr}{k_1^2+nl_1}dv+\frac{2k_1^3+3nk_1l_1}{2nR_1^2}(d\psi'+n\cos\theta d\phi)\right], 
\label{NHgauge}
\end{eqnarray}
where we have defined 
\begin{eqnarray}
&&R_1^2:=k_1^2+nl_1,\label{sec:R1ineq}\\
&&R_2^2:=\frac{l_1^2(3k_1^2+4nl_1)}{R_1^4}.\label{sec:R2ineq}
\end{eqnarray}
This is locally isometric to the near-horizon geometry of the BMPV black hole~\cite{Chamseddine:1996pi,Gauntlett:1998fz}.\footnote{In spite of this, one can verify that the metric (\ref{NHmetric}) with (\ref{NHgauge}) preserves the maximal amount of supersymmetry, as in the near-horizon geometry of the BMPV black hole. This is quite nontrivial because the discrete identification in general breaks supersymmetry. The primary reason for this is that $\partial/\partial\psi'$ is a symmetry of two independent Killing spinors. This can be demonstrated by a direct integration of the Killing spinor equations~(\ref{KS}). }
In order to remove CTCs around the horizon, one requires 
$R_1^2>0$ and $R_2^2>0$, which simply amounts to the inequality
\begin{eqnarray}
\label{l1_ineq}
l_1>- \frac{3k_1^2}{4n} \,. \label{eq:R2^2-positive}
\end{eqnarray}

The cross section of the event horizon
can be extracted by $v={\rm const.}$ and $r=0$ in (\ref{NHmetric}), giving rise to 
\begin{eqnarray}
ds^2_{\rm H}=\frac{R_2^2}4 (d \psi'+n \cos\theta d\phi)^2+R_1^2(d\theta^2+\sin^2\theta d\phi^2) \,, 
\end{eqnarray}
which precisely recovers the squashed metric of the lens space $S^3/\mathbb Z_n$ given in (\ref{lens}).




\subsection{Bubbles ${\bm r}={\bm r}_i$ ($n=2,...,n$)}

There exists apparent divergence in the metric of the Gibbons-Hawking space at the points  
${\bm r}={\bm r}_i$ ($n=2,...,n$). We shall impose the boundary conditions at each point 
${\bm r}={\bm r}_i$ ($n=2,...,n$) that this corresponds to a smooth point analogous to the origin of Minkowski spacetime, 
rather than horizons.
To demonstrate this, let us choose the coordinates $x^i$ on ${\mathbb E}^3$ of the Gibbons-Hawking space so that the $i$-th point ${\bm r}={\bm r}_i$ ($i\not=1$) is an origin of ${\mathbb E}^3$. 
Near the origin ${\bm r}=0$, the four harmonic functions $H$, $K$, $L$ and $M$ behave as  
\begin{eqnarray}
H&\simeq&-\frac{1}{r}+\sum_{j(\not=i)}\frac{h_j}{|z_{ji}|},\quad K\simeq\frac{k_i}{r}+\sum_{j(\not=i)}\frac{k_j}{|z_{ji}|},\quad \\
L&\simeq&\frac{l_i}{r}+1+\sum_{j(\not=i)}\frac{l_j}{|z_{ji}|},\quad  M\simeq\frac{m_i}{r}+m_0+\sum_{j(\not=i)}\frac{m_j}{|z_{ji}|}, 
\end{eqnarray}
yielding 
\begin{eqnarray}
f^{-1}&\simeq& -\frac{k_i^2-l_i}{r}+c_1,\qquad 
\omega_\psi\simeq \frac{k_i^3-\frac{3}{2}k_il_i+m_i}{r}+c_2,
\end{eqnarray}
where the constants $c_1$ and $c_2$ are defined by
\begin{eqnarray}
c_1&:=&1+\sum_{j(\not=i)}\frac{1}{|z_{ji}|}(l_j-2k_ik_j-k_i^2h_j),\\
c_2&:=&m_0-\frac{3}{2}k_i+\sum_{j(\not =i)}\frac{1}{|z_{ji}|}[3k_i^2k_j+2k_i^3h_j-\frac{3}{2}(k_il_j+l_ik_j+k_il_ih_j)+m_j].\label{eq:c2}
\end{eqnarray}
The 1-forms $\tilde \omega_i$ and $\hat\omega_{ij}$ are approximated by 
\begin{eqnarray}
&&\tilde \omega_i\simeq \cos\theta d\phi,\qquad 
\tilde \omega_j\simeq -\frac{z_{ji}}{|z_{ji}|} d\phi\ (j\not=i),\\
&&\hat \omega_{ij}\simeq -\frac{\cos\theta}{|z_{ji}|} d\phi\ (i\not= j) ,\qquad \hat \omega_{kj}\simeq \frac{z_{ji}z_{ki}}{|z_{ji}z_{ki}|z_{jk}} d\phi\ (k\not =j, k,j\not=i), 
\end{eqnarray}
giving rise to 
\begin{eqnarray}
\chi&\simeq& \left(-\cos\theta+\chi_{(0)}\right)d\phi\,, \qquad 
\hat\omega\simeq(\hat \omega_{(1)}\cos\theta+\hat\omega_{(0)})d\phi \,,
\end{eqnarray}
where 
\begin{eqnarray}
\chi_{(0)} & := &- \sum_{j(\not=i)}\frac{h_jz_{ji}}{|z_{ji}|} \,,\\
\hat \omega_{(0)}& := &  \sum_{k,j(\not=i,k\not=j)}\left(h_km_j+\frac{3}{2}k_kl_j\right)\frac{z_{ji}z_{ki}}{|z_{ji}z_{ki}|z_{jk}}+\sum_{j(\not= i)}\left(m_0h_j+\frac{3}{2}k_j\right)\frac{z_{ji}}{|z_{ji}|}+c\,, \\
\hat \omega_{(1)}& := & -\sum_{j(\not=i)}\left(h_im_j-h_jm_i+\frac{3}{2}(k_il_j-k_jl_i)\right)\frac{1}{|z_{ji}|}-\left(m_0h_i+\frac{3}{2}k_i\right)\,.
\end{eqnarray}
One therefore obtains the asymptotic behavior of the metric around the $i$-th point  as
{\small
\begin{eqnarray}
ds^2&\simeq& -\left( -\frac{k_i^2-l_i}{r}+c_1\right)^{-2}\left[dt+\left(\frac{k_i^3-\frac{3}{2}k_il_i+m_i}{r}+c_2\right)\left\{d\psi+(-\cos\theta+\chi_{(0)})d\phi \right\} +(\hat \omega_{(1)}\cos\theta+\hat\omega_{(0)})d\phi \right]^2\nonumber\\
&-&\left( -\frac{k_i^2-l_i}{r}+c_1\right)r\left[\left\{d\psi+(-\cos\theta+\chi_{(0)})d\phi \right\}^2+\frac{dr^2}{r^2}+d\theta^2+\sin^2\theta d\phi^2\right].
\end{eqnarray}
}

As a minimal requirement, we impose the following conditions on the parameters $(k_i,l_i,m_i)\ (i=2,...,n)$:
\begin{eqnarray}
l_i&=&k_i^2,\label{eq:condition1}\\
m_i&=&\frac{1}{2}k_i^3\label{eq:condition2},
\end{eqnarray}
which implies 
\begin{eqnarray}
k_i^3-\frac{3}{2}k_il_i+m_i=0,\qquad c_2=\hat\omega_{(1)}.
\end{eqnarray}
Hence, (\ref{eq:condition1}) and (\ref{eq:condition2}) are sufficient to get rid of 
$1/r$ divergence in $g_{\mu\nu}$  and ensure that no curvature singularities appear 
in the domain of outer communications. 
Converting to the new coordinates $(\rho,\psi', \phi')$ by
\begin{eqnarray}
\rho=2\sqrt{-c_1r}, \qquad  \psi'=\psi+\chi_{(0)}\phi, \qquad \phi'=\phi \,, 
\end{eqnarray}
one verifies that the metric near  ${\bm r}={\bm r}_i$ reduces to
\begin{eqnarray}
ds^2\simeq -c_1^{-2} d[t+c_2\psi'+\hat \omega_{(0)}\phi']^2+\left[d\rho^2+\frac{\rho^2}{4}\left\{(d\psi'-\cos\theta d\phi')^2+d\theta^2+\sin^2\theta d\phi'{}^2\right\}\right] \,. 
\label{nutmetric}
\end{eqnarray}
To ensure the metric the Lorentzian signature, one needs
\begin{eqnarray}
c_1<0 \,. \label{eq:c1ineq}
\end{eqnarray}
Although (\ref{nutmetric}) is locally isometric to the flat space metric, 
$\partial/\partial \psi'=\partial/\partial \psi$ necessarily becomes timelike and CTCs appear, because
we are now focusing on the region $\rho \simeq 0$. 
To remove this causal violation around each ${\bm r}_i$, it suffices to 
impose $c_2=0$ at ${\bm r}={\bm r}_i$ ($i=2,...,n$), which in the present case rehashes to 
\begin{eqnarray}
m_0-\frac{3}{2}k_i+\sum_{j(\not =i)}\frac{1}{|z_{ji}|}\left[3k_i^2k_j+2k_i^3h_j-\frac{3}{2}(k_il_j+l_ik_j+k_il_ih_j)+m_j\right]=0.
\label{eq:condition3}
\end{eqnarray}
These conditions are obtained by requiring $\omega_\psi=0$ at ${\bm r}={\bm r}_i$ ($i=2,...,n$) and referred to as ``bubble equations'' in refs~\cite{Bena:2005va,Bena:2007kg}. 
These bubbles ${\bm r}={\bm r}_i$ describe the timelike and regular surfaces. 
The bubble equations account for the delicate balance between the 
gravitational attraction and the repulsion by flux through the cycles.  

There still remains a possibility of CTCs associated with a vector field with closed orbits
$\partial/\partial \phi'=\partial/\partial \phi-\chi_{(0)} \partial/\partial \psi$ remains spacelike (note that $\chi_{(0)}$ is integer). 
Quite amazingly, 
$\hat\omega_{(0)}=0$ is automatically satisfied for all $i=2,...,n$,  if we impose (\ref{eq:condition3}). 
We relegate the proof of this in Appendix \ref{B}. 
It follows that  no causal violation occurs around each bubble if one demands (\ref{eq:condition3}).

As a consistency check, 
the conditions (\ref{eq:c1ineq}) and (\ref{eq:condition3})  
reduce respectively  to the equations (21) and (11) in~\cite{Kunduri:2014kja} in 
the $n=2$ case. One can also verify that the metric and the Maxwell field are smooth 
at each bubble.

\subsection{Axis}

The $z$-axis of ${\mathbb E}^3$ (i.e., $x=y=0$) in the Gibbons-Hawking space splits up into  the $(n+1)$ intervals: $I_-=\{(x,y,z)|x=y=0,  z<z_1\}$, $I_i=\{(x,y,z)|x=y=0,z_i<z<z_{i+1}\}\ (i=1,...,n-1)$ and $I_+=\{(x,y,z)|x=y=0,z>z_n\}$. On the $z$-axis, the $1$-forms $\hat\omega_{ij}$ and $\tilde\omega_i$  are, respectively, simplified to 
\begin{eqnarray}
\hat \omega_{ij}=\frac{(z-z_i)(z-z_j)}{z_{ji}|z-z_i||z-z_j|}d\phi,\qquad \tilde\omega_i=\frac{z-z_i}{|z-z_i|}d\phi.
\end{eqnarray}
In particular, on $ I_\pm$, $\hat\omega_{ij}$ and $\tilde\omega_i$ become, respectively,
\begin{eqnarray}
\hat \omega_{ij}=\frac{1}{z_{ji}}d\phi,\qquad \tilde\omega_i=\pm d\phi.
\end{eqnarray}
Hence,  on $I_\pm$, $\hat\omega$ vanishes since
\begin{eqnarray}
\hat\omega&=&\sum_{k,j(k\not=j)}\left(h_km_j+\frac{3}{2}k_kl_j\right)\hat\omega_{kj}-\sum_{j}\left(m_0h_j+\frac{3}{2}k_j\right)\hat\omega_j+c d\phi\notag\\
&=&\sum_{k,j(k\not=j)}\left(h_km_j+\frac{3}{2}k_kl_j\right)\frac{d\phi}{z_{jk}}\mp\sum_{j}\left(m_0h_j+\frac{3}{2}k_j\right)d\phi -\sum_{k,j(k\not=j)}\left(h_km_j+\frac{3}{2}k_kl_j\right)\frac{d\phi}{z_{jk}} \notag\\
&=&\mp\sum_{j}\left(m_0h_j+\frac{3}{2}k_j\right)d\phi \notag\\
&=&\mp\left( m_0+\frac{3}{2}\sum_jk_j \right) d\phi\notag\\
&=&0,
\end{eqnarray}
where we have used Eq.~(\ref{eq:m0}) at the last equality. 

On the interval $I_1$, one computes
\begin{eqnarray}
\hat\omega_{\phi}-\hat\omega_{(0)}&=-&\sum_{2\le j} \frac{h_1m_j-h_jm_1+\frac{3}{2}(k_1l_j-k_jl_1)}{z_{j1}}\frac{(z_j-z)}{|z_j-z|}-\left(m_0h_1+\frac{3}{2}k_1\right)\notag \\
                                                  &=&-\sum_{2\le j}\frac{nk_j^3+3k_1k_j^2-3k_jl_1}{2|z_{j1}|}-\left(nm_0+\frac{3}{2}k_1\right)\notag \\
                                                  &=&\sum_{2\le j}\left(m_0-\frac{3}{2}k_j\right)-\left(nm_0+\frac{3}{2}k_1\right)\notag \\
                                                  &=&-\left(m_0+\sum_j\frac{3}{2}k_j\right)\notag \\
                                                  &=&0,
\end{eqnarray}
where we have used Eq.~(\ref{eq:sumc2=02}). Finally  for $z\in I_i\ (i=2,...,n-1)$
we get
\begin{eqnarray}
\hat\omega_{\phi}-\hat\omega_{(0)}&=&\sum_{j(\not=i)}\frac{h_im_j-h_jm_i+\frac{3}{2}(k_il_j-k_jl_i)}{z_{ji}}\frac{(z_i-z)(z_j-z)}{|z_i-z||z_j-z|}+\left(m_0h_i+\frac{3}{2}k_i\right)\frac{z_i-z}{|z_i-z|}\notag \\
&=&\frac{h_im_1-h_1m_i+\frac{3}{2}(k_il_1-k_1l_i)}{z_{1i}}-\sum_{2\le j(\not=i)}\frac{h_1m_j-h_jm_1+\frac{3}{2}(k_1l_j-k_jl_1)}{z_{ji}}\frac{(z_j-z)}{|z_j-z|}\notag \\
&&-\left(m_0h_i+\frac{3}{2}k_i\right) \notag \\
                    &=&\frac{-nk_i^3+3k_il_1-3k_1k_i^2}{2z_{1i}}-\sum_{2\le j(\not=i)}\frac{(k_i-k_j)^3}{2z_{ji}}\frac{(z_j-z)}{|z_j-z|}+\left(m_0-\frac{3}{2}k_i\right)\notag  \\
                    &=&c_2\notag \\
                    &=&0,                 
\end{eqnarray}
where we have used Eq.~(\ref{eq:c2=0}). 
Since $\hat\omega_{(0)}=0$ (see Appendix \ref{B}), 
one concludes  $\hat\omega_\phi=0$ (i.e., $\hat\omega=0$) on $I_i$ for $i=1,...,n-1$.
It therefore turns out that  $\hat\omega=0$ holds at each interval. 
This proves that no Dirac-Misner string pathologies happen throughout the spacetime. 
As argued in \cite{Bena:2007kg,Gibbons:2013tqa}, the absence of the Dirac-Misner string among each bubble is a direct consequence of the bubble equations (\ref{eq:condition3}). 
However,  $\hat\omega=0$ at $I_1$ seems quite nontrivial.

\medskip

Let us next move onto the discussion of the issue of orbifold singularities. 
On $I_\pm$ we get 
\begin{eqnarray}
\chi&=&\pm d\phi,
\end{eqnarray} 
whereas on $I_i$ we have 
\begin{eqnarray}
\chi
      &=&\left(n\frac{z-z_1}{|z-z_1|}-\sum_{2\le j\le i}\frac{z-z_j}{|z-z_j|}-\sum_{i+1\le j\le n-1}\frac{z-z_j}{|z-z_j|}\right)d\phi \notag \\
      &=&\left(2n-2i+1\right)d\phi.
\end{eqnarray}
Accordingly, 
the two-dimensional $(\phi,\psi)$-part of the metric on the intervals $I_\pm$ and $I_i$
can be written in the form
\begin{eqnarray}
ds^2_2=(-f^2\omega_\psi+f^{-1}H^{-1})(d\psi+\chi_\phi d\phi)^2. \label{eq:axis}
\end{eqnarray}
In the analysis of orbifold singularities, it is more advantageous to work in 
the coordinate basis vectors $(\partial_{\phi_1},\partial_{\phi_2})$ of $2\pi$ periodicity, 
instead of $(\partial_\phi,\partial_\psi)$, where these coordinates are defined by $\phi_1:=(\psi+\phi)/2$ and $\phi_2:=(\psi-\phi)/2$.  
From~(\ref{eq:axis}), one sees that  the Killing vector $v:=\partial_\phi-\chi_\phi\partial_\psi$ vanishes on each interval. More precisely, one sees that
\begin{itemize}
\item on the interval $I_+$, the Killing vector $v_+:=\partial_\phi-\partial_\psi=(1,0)$ vanishes,
\item on each interval $I_i$ ($i=1,...,n-1$), the Killing vector $v_i:=\partial_\phi-(2n-2i+1)\partial_\psi=(1,i-n)$ vanishes,
\item on the interval $I_-$, the Killing vector $v_-:=\partial_\phi+\partial_\psi=(1,1)$ vanishes. 
\end{itemize}
From these, we can observe that the Killing vectors $v_\pm,\ v_i$ on the intervals satisfy 
\begin{eqnarray}
{\rm det}\ (v_+^T,v_{n-1}^T)=-1,\qquad {\rm det}\ (v_{i}^T,v_{i-1}^T)=-1, 
\label{noorbifold}
\end{eqnarray}
with 
\begin{eqnarray}
{\rm det}\ (v_1^T,v_{-}^T)=n.
\label{lens_cond}
\end{eqnarray}
Eq. (\ref{noorbifold}) assures that the metric smoothly joints at the end points $z=z_i\ (1\le i\le n)$ of the intervals~\cite{Hollands:2007aj}, which means that there exist no orbifold singularities at adjacent 
intervals. 
Furthermore, eq. (\ref{lens_cond}) illustrates that the horizon cross section is topologically the lens space $L(n,1)=S^3/{\mathbb Z}_n$.

\subsection{Summary of boundary conditions}

The regularity of the metric at each boundary has requested us to impose 
the conditions (\ref{eq:l0}), (\ref{eq:c}), (\ref{eq:m0}), (\ref{eq:condition1}), (\ref{eq:condition2}) and (\ref{eq:condition3}). Combined with $m_1=z_1=0$, this reduces the independent parameters of the
solution from $4n+1$ to $n+1$. 
Moreover, these parameters are subject to the constraints 
(\ref{l1_ineq}) and (\ref{eq:c1ineq}). The latter condition boils down to 
\begin{eqnarray}
\frac{nk_i^2+2k_i k_1-l_1}{z_{i1}}> 1+\sum_{j (\ne i) \ge 2} \frac{(k_i-k_j)^2}{|z_{ji}|} \,, \qquad i=2,...,n. 
\label{eq:negative_c1}
\end{eqnarray}




\section{Physical properties}
\label{sec:analysis}
As appropriate boundary conditions are prescribed in the last section, 
let us now investigate several physical properties of the solution. 

\subsection{Conserved quantities}

Let us discuss conserved quantities of the black lens solution. As shown in section~\ref{sec:bc_infinity}, the spacetime is asymptotically flat, which enables us to ADM mass 
the ADM mass and two ADM angular momenta as
\begin{eqnarray}
M&=&\frac{\sqrt{3}}{2}Q=3\pi\left[\left(\sum_{i}k_i\right)^2+\sum_{i}l_i\right],\\
J_\psi&=&4\pi \left[ \left(\sum_{i}k_i\right)^3
+\frac{1}{2}\sum_{i=2}^nk_i^3+\frac{3}{2}\left(\sum_{i}k_i\right)\left(l_1+\sum_{i=2}^nk_i^2\right)\right],\\
J_\phi&=&6\pi\left[  \left(\sum_{i}k_i\right) \left(
\sum_{j=2}^nz_j\right)+\left(\sum_{i =2}^nk_iz_i\right)  \right], 
\end{eqnarray}
where $Q$ is the electric charge normalized by (note that the Chern-Simons term fail to contribute) 
\begin{eqnarray}
Q=\int _{\cal S} \star F \,. 
\end{eqnarray}
It follows that the Bogomol'ny bound is saturated~\cite{Gibbons:1993xt}. 
In appearance, the positivity of the ADM mass is not obvious from the above expressions. 
Nevertheless, one can establish the positivity of the mass by 
using the relations~(\ref{sec:R1ineq}) and (\ref{eq:condition1}) as
\begin{eqnarray}
\frac{M}{3\pi}&=&\left(\sum_{i}k_i\right)^2+\sum_{i}l_i \notag\\
                     &>&k_1^2+\left(\sum_{i\not=1}k_i\right)^2+2\left(\sum_{i\not=1}k_i\right)k_1-\frac{1}{n}k_1^2+\sum_{i\not=1}l_i\notag\\
                     &=&\frac{n-1}{n}\left(k_1+\frac{n}{n-1}\sum_{i\not=1}k_i\right)^2-\frac{1}{n-1}\left(\sum_{i\not=1}k_i\right)^2+\sum_{i\not=1}k_i^2\notag\\
                      &=&\frac{n-1}{n}\left(k_1+\frac{n}{n-1}\sum_{i\not=1}k_i\right)^2+\frac{1}{n-1}\left[(n-2)\sum_{i\not=1}k_i^2-2\sum_{i,j(2\le i<j)}k_ik_j\right]\notag\\
                     &=&\frac{n-1}{n}\left(k_1+\frac{n}{n-1}\sum_{i\not=1}k_i\right)^2+\frac{1}{n-1}\sum_{i,j(2\le i<j)}(k_i-k_j)^2>0, 
\end{eqnarray}
where we have used the inequality~(\ref{sec:R1ineq}).

The surface gravity and the angular velocities of the horizon vanish, as expected
for supersymmetric black objects in the asymptotically flat spacetime~\cite{Gauntlett:1998fz}.
The area of the horizon is nonvanishing  and reads from (\ref{NHmetric}) as
\begin{eqnarray}
{\rm Area}=8\pi^2 R_1^2 R_2 . 
\end{eqnarray}

The  interval $I_1$ has a disc topology and the $(n-2)$ intervals $I_i$ ($i=2,...,n-1$) is a two-dimensional sphere, respectively. The magnetic fluxes through $I_i$ are defined as 
\begin{eqnarray}
q[I_i]:=\frac{1}{4\pi}\int_{I_i}F\,.
\end{eqnarray}
Since the Maxwell gauge field $A_\mu$ is smooth 
at the horizon, bubbles and critical surfaces, one can evaluate these fluxes 
as $q[I_i]=[-A_\psi]^{z=z_{i+1}}_{z=z_i}$, 
which are computed to give\footnote{Perhaps, ref.~\cite{Kunduri:2014kja} misses the contribution coming from the horizon. }
\begin{eqnarray}
q[I_1]=\frac{\sqrt{3}}{2}\left[
\frac{k_1l_1}{2(k_1^2+nl_1)}
-k_2
\right]\,, \qquad 
q[I_i]=\frac{\sqrt{3}}{2}(k_i-k_{i+1})~~ (i=2,... n-1).
\label{mag_fluxes}
\end{eqnarray}
An alternative definition of magnetic flux proposed in~\cite{Bena:2007kg}  is 
\begin{eqnarray}
\tilde q[I_j] \equiv -\frac{1}{4\pi \sqrt 3} \int _{I_j} G^+ \,. 
\end{eqnarray}
This definition does not apply in the present setting, because $G^+$ diverges 
at the critical surfaces at which $H=0$. 
The interval $I_1$ corresponds to the disk topology surface ending on the horizon, whereas the 
intervals $I_i$ ($i=2,...,n$) represents 2-cycles (or bubbles) outside the horizon.

When $k_2=k_1l_1/[2(k_1^2+nl_1)]$, $k_i=k_{i+1}\ (i=2,...,n-1)$, all magnetic fluxes $q[I_i]\ (i=1,...,n-1)$ vanish. 
Under this parameter-setting, the condition~(\ref{eq:negative_c1}) can be simply written as 
\begin{eqnarray}
-\frac{nl_1^2(3k_1^2+4nl_1)}{4(k_1^2+nl_1)^2}>z_{i1}, \qquad i=2,...,n.
\end{eqnarray}
However, it turns out that these inequalities cannot be satisfied, since the left-hand side is non-positive by~(\ref{eq:R2^2-positive}) while the right-hand side is positive by our assumption. 
As for the supersymmetric solutions which we have obtained, one may therefore interpret that 
the horizon of the lens space topology $L(n,1)=S^3/{\mathbb Z}_n$ must be supported by the magnetic fluxes.

When 
\begin{eqnarray}
k_1=-\sum_{i=2}^nk_i-\frac{\sum_{i=2}^nk_iz_i}{\sum_{i=2}^nz_i},\label{eq:J=0}
\end{eqnarray}
the angular momentum  $J_\phi$ vanishes, which implies that the black lens has equal angular momenta $J_{\phi_1}=J_{\phi_2}$ as  the BMPV black hole does.  
For $n=2$, the inequality~(\ref{eq:negative_c1}) is never  satisfied in the choice of the parameters~(\ref{eq:J=0}),  which means that the black lens with equal angular momenta cannot be realized~\cite{Kunduri:2014kja}.  
For $n\ge 3$, it seems impracticable to show this claim analytically. 
As far as we checked numerically for $n=3$, we find no parameter regions under which the 
configuration of equal angular momenta is realized.  
We expect that this situation does not change  for $n\ge 4$, and 
our family of black lenses does not admit equal angular momenta.

\subsection{No CTCs}

We wish to impose that the domain of outer communication in the five-dimensional spacetime remains Lorenzian without CTCs. 
This amounts to
\begin{eqnarray}
g_{\theta\theta}>0, \qquad g_{\psi\psi}>0 \,, \qquad 
g_{\psi\psi}g_{\phi\phi}-g_{\psi\phi}^2>0 \,. 
\end{eqnarray}
These conditions ensure that $t$ is a global timelike coordinate and the spacetime
is stably causal. 
Explicitly, these conditions boil down to 
\begin{align}
\label{}
D_1:& = K^2+HL >0 \,, \\ 
D_2:& = \frac 34 K^2L^2-2K^3 M-3 HKLM+HL^3-H^2 M^2 \,, \\
D_3:& =  D_2 r^2\sin^2\theta-\hat \omega_\phi^2 \,,  
\end{align} 
It is considerably elaborate to check their positivity. For instance, one finds 
\begin{align}
K^2+HL=&\frac{k_1^2+nl_1}{r_1^2}+\frac{n}{r_1}+\sum_{i\ge 2}\frac{1}{r_{i}}\left[
\frac{nk_{i}^2+2k_1k_{i}-l_1}{r_1}-1-\sum_{j\ge 2}\frac{(k_{i}-k_{j})^2}{2r_{j}}
\right]\,, \notag \\
\ge &\frac{k_1^2+nl_1}{r_1^2}+\frac{1}{r_1}+\sum_{j\ge 2}\left[
\left\{
\frac{|z_{1j}|}{r_1r_j}-\frac{1}{r_j}+\frac{1}{r_1}
\right\}+\sum_{k(\ne j)\ge 2}(k_j-k_k)^2\left\{
\frac{|z_{1j}|}{r_1|z_{kj}|}-\frac{1}{2r_k}
\right\}
\right]\,, 
\end{align}
where we have used (\ref{eq:negative_c1}). 
Due to the triangle inequality $r_j+|z_{1j}|\ge r_1$, we find that 
the first term in the summation of the right-hand side is non-negative, while the 
second term is rewritten into
\begin{align}
 \sum_{j\ge 2}\sum_{k(\ne j)\ge 2}(k_j-k_k)^2\left\{
\frac{|z_{1j}|}{r_1|z_{kj}|}-\frac{1}{2r_k}
\right\}=& \frac{1}{2}\sum_{j\ge 2}\sum_{k(\ne j)\ge 2}\frac{(k_j-k_k)^2}{2r_1r_jr_k|z_{kj}|}
\left[2r_jr_k(|z_{1j}|+|z_{1k}|)-r_1|z_{kj}|(r_j+r_k)\right]\,
\notag \\
\ge &\sum_{j\ge 2}\sum_{k(\ne j)\ge 2}\frac{(k_j-k_k)^2}{4r_1r_jr_k}\{2r_jr_k-r_1(r_j+r_k)\}
\end{align}
where we used $|z_{1j}|+|z_{1k}|\ge |z_{kj}|$. 
This term vanishes for $n=2$, whereas its positivity is unclear for $n>2$. 

\begin{figure}[t]
\label{fig:CTC}
\begin{center}
\includegraphics[width=15cm]{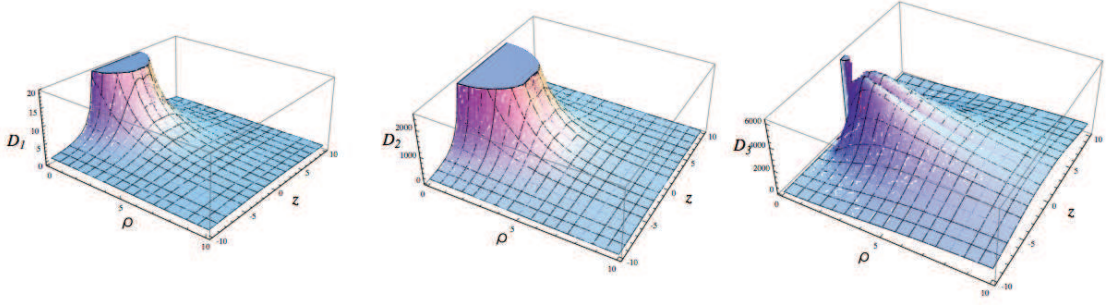}
\caption{Plots of $D_i$'s against ($\rho=\sqrt{x^2+y^2}, z$) 
for $n=4$, $z_1=0$, $k_1=l_1=z_2=1$, $z_2=1$, $z_4=3$ with 
$k_2\simeq -2.36$, $k_3\simeq -2.88$, $k_4\simeq -3.27$. No naked time machines 
appear.}
\end{center}
\end{figure}

Nevertheless, we made substantial numerical scans for the positivity of $D_i$'s and found that 
there appear no causal violations in the domain of outer communications (see Fig.~\ref{fig:CTC}).   
We expect that (\ref{l1_ineq}) and  (\ref{eq:negative_c1}) are sufficient to remove causal pathologies on and outside the horizon.

\subsection{Critical surfaces}

A key issue for the construction of the black lens is the contrived choice
of harmonic functions (\ref{Hdef})--(\ref{Ldef}). One immediate notable feature is that $H$ is negative
around ${\bm r}={\bm r}_i$ ($i=2,...,n$), giving rise to the ($-,-,-,-$) signature of the Gibbons-Hawking base space. 
This is not problematic as long as $f^{-1}H$ remains positive. 
There appear ``critical surfaces'' \cite{Bena:2007kg} at which $f=0$ corresponding to $H=0$.  
From the five-dimensional point of view, this is called an ``evanescent ergosurface'' \cite{Gibbons:2013tqa,Niehoff:2016gbi}. 
This is not problematic since these surfaces exist also in ${\rm AdS}_3\times S^2$~\cite{Gibbons:2013tqa}.

The existence of critical surfaces provokes a striking impact on the uniqueness theorems of supersymmetric black hole. 
In the original proof \cite{Reall:2002bh}, it has been assumed that the supersymmetric Killing field 
is strictly timelike outside the horizon. Recently it has been pointed out in \cite{Kunduri:2014iga}
that these critical surfaces can get around the uniqueness theorems. 
Since the topologically nontrivial cycles run between two point sources for the Gibbons-Hawking harmonic function, this way of avoiding uniqueness theorems does not happen in four dimensions.

At these regularity surfaces, one must impose the regularity condition 
$K \ne 0$ when $H=0$.  Namely, if there exist points $z=z_c$ on the axis 
such that 
\begin{eqnarray}
H=\sum_{i}\frac{h_i}{|z_c-z_i|}=0,\qquad 
K=\sum_{i}\frac{k_i}{|z_c-z_i|}=0, 
\label{HKsingCS}
\end{eqnarray}
these critical surfaces are singular. In $n=2$ case, this does not occur 
by the restriction (\ref{eq:negative_c1}), whereas the $n\ge 3$ case 
seems nontrivial, although we have not found the singular behavior numerically. 
If these surfaces were singular, (\ref{HKsingCS}) gives rise to 
\begin{eqnarray}
\sum_{i\ge 2}\frac{nk_i+k_1}{|z_c-z_i|}=0. 
\end{eqnarray}
Hence, a sufficient condition to avoid the singularity at critical surfaces 
is 
\begin{eqnarray}
k_1+nk_i \ne 0   \,, \qquad (2\le i \le n) \,. 
\label{eq:additional}
\end{eqnarray}




\section{Summary}
\label{sec:discuss}

In this work,  we have constructed an asymptotically flat supersymmetric black lens solution in the bosonic sector of the five-dimensional minimal supergravity, whose horizon topology is a lens space of $L(n,1)=S^3/\mathbb{Z}_n$ ($n=1,2,...$). 
This represents a generalization of the solution found in~\cite{Kunduri:2014kja}. 
We also computed the conserved charges including  the (positive and BPS-saturating) mass, two angular momenta, and $(n-1)$ magnetic fluxes, among which $n+1$ quantities are independent.

As is concerned with the black lens solution which we have obtained in this work,  there exists no limit  such that all the magnetic fluxes vanish. 
Therefore, as for the supersymmetric solutions, the existence of the magnetic fluxes seem to play an essential role in supporting the horizon of the black lens. In general, however, it is not clear whether one necessarily needs such magnetic fluxes in order to construct a black lens. 

It appears straightforward to generalize the present work to into ${\rm U}(1)^3$ supergravity 
and uplift the solution into 11 dimensions~\cite{Elvang:2004ds}. Performing the Kaluza-Klein reduction and the subsequent T-dualities, the solution can be converted into the D1-D5-P system. We expect that a decoupling limit can be taken as in the $n=2$ case, which would enable us to exploit Cardy formula to reproduce the Bekenstein-Hawking entropy.

In this line of research, one of the most exciting generalizations might be to look for a non-supersymmetric black lens. 
Such a black lens, if they exist, in particular a vacuum solution, must differ considerably from the solutions presented here, since 
we have mede use of the flux threading the 2-cycles to prevent the collapse, while in the vacuum case this does not occur~\cite{Gibbons:2013tqa}.




\acknowledgments
This work was partially supported by the Grant-in-Aid for Young Scientists (B) (No.~26800120) from Japan Society for the Promotion of Science (S.T.), and by MEXT Grant-in-Aid for Scientific Research on Innovative Areas``New Developments in Astrophysics Through Multi-Messenger Observations of Gravitational Wave Sources" (Grant Number A05 24103006) (M.N).




\appendix




\section{Proof of $\hat\omega_{(0)}=0$}
\label{B}

At each bubble, no causal violation requires that 
$\hat\omega_{(0)}=0$ holds for $i=2,...,n$. To prove this is the main result of the present appendix. 
Let us note that from (\ref{eq:condition1}) and (\ref{eq:condition2}),  
the bubble equations  (\ref{eq:condition3}) can be written as 
\begin{eqnarray}
0&=&m_0-\frac{3}{2}k_il_0+\sum_{j(\not =i)}\frac{1}{|z_{ji}|}[3k_i^2k_j+2k_i^3h_j-\frac{3}{2}(k_il_j+l_ik_j+k_il_ih_j)+m_j]\notag \\
 &=&m_0-\frac{3}{2}k_i-\frac{nk_i^3+3k_1k_i^2-3l_1k_i}{2z_{1i}}+\sum_{2 \le j(\not =i)}\frac{(k_j-k_i)^3}{2|z_{ji}|}.\label{eq:c2=0}
 \end{eqnarray}
 The summation  of (\ref{eq:c2=0}) for $i=2,..., n$ gives
\begin{eqnarray}
 0&=&\sum_{2\le j}\left[m_0-\frac{3}{2}k_j-\frac{nk_j^3+3k_1k_j^2-3l_1k_j}{2z_{1j}}+\sum_{2 \le k(\not =i)}\frac{(k_k-k_j)^3}{2|z_{kj}|}\right] 
 \notag\\ 
&=&\sum_{2\le j}\left(m_0-\frac{3}{2}k_j\right)-\sum_{2\le j }\frac{nk_j^3+3k_1k_j^2-3l_1k_j}{2z_{1j}},\label{eq:sumc2=02}
\end{eqnarray}
where the last term in the first line vanishes by the antisymmetry for $k$ and $j$.
From Eqs.~(\ref{eq:condition1}) and (\ref{eq:condition2}), $\hat\omega_{(0)}$ is written as
\begin{eqnarray}
\hat\omega_{(0)}&=&\sum_{k,j(k,j\not=i,k\not=j)}\left(h_km_j+\frac{3}{2}k_kl_j\right)\frac{z_{ji}z_{ki}}{|z_{ji}z_{ki}|z_{jk}}+\sum_{j(\not=i)}\left( m_0h_j+\frac{3}{2}k_j\right)\frac{z_{ji}}{|z_{ji}|}\nonumber \\
&&-\sum_{k,j(j\not=k)}\frac{h_km_j+\frac{3}{2}k_kl_j}{z_{jk}}\notag \\
&=&-\sum_{2\le j(\not=i)}\frac{nk_j^3+3k_1k_j^2-3k_jl_1}{2z_{j1}}\frac{z_{ji}}{|z_{ji}|}+\sum_{2\le k,j (k,j \not=i,k\not=j)}\frac{-k_j^3+3k_kk_j^2}{2z_{jk}}\frac{z_{ji}z_{ki}}{|z_{ji}z_{ki}|}\nonumber\\
&&-\left(nm_0+\frac{3}{2}k_1\right)+\sum_{2\le j(\not=i)}\left( -m_0+\frac{3}{2}k_j\right)\frac{z_{ji}}{|z_{ji}|}\notag \\
&&-\sum_{2\le j}\frac{nk_j^3+3k_1k_j^2-3l_1k_j}{2z_{j1}}-\sum_{2\le k,j(k\not=j)}\frac{-k_k^3+3k_k^2k_j}{2z_{kj}}. \label{eq:omega}
\end{eqnarray}
The third, fifth and sixth terms of the right-hand side of (\ref{eq:omega}) are combined into
\begin{eqnarray}
&&-\left(nm_0+\frac{3}{2}k_1\right)-\sum_{2\le j}\frac{nk_j^3+3k_1k_j^2-3l_1k_j}{2z_{j1}}-\sum_{2\le k,j(k\not=j)}\frac{-k_k^3+3k_k^2k_j}{2z_{kj}}\notag \\
&&~~=-\left(nm_0+\frac{3}{2}k_1\right)+\sum_{2\le j}\left(m_0-\frac{3}{2}k_j\right)+\sum_{2\le k,j (k\not=j)}\frac{(k_k-k_j)^3}{4z_{kj}}\notag \\
&&~~=\sum_{2\le j,k(k\not=j)}\frac{(k_k-k_j)^3}{4z_{kj}},\label{eq:356}
\end{eqnarray}
where we have used Eq.~(\ref{eq:c2=0}). 
Next, the summation of the first, second and fourth terms  on the right-hand
side of (\ref{eq:omega}) reduces to 
\begin{eqnarray}
&-&\sum_{2\le j(\not=i)}\frac{nk_j^3+3k_1k_j^2-3k_jl_1}{2z_{j1}}\frac{z_{ji}}{|z_{ji}|}+\sum_{2\le k,j (k,j\not=i,k\not=j)}\frac{-k_j^3+3k_kk_j^2}{2z_{jk}}\frac{z_{ji}z_{ki}}{|z_{ji}z_{ki}|}\notag \\
&&+\sum_{2\le j(\not=i)}\left( -m_0+\frac{3}{2}k_j\right)\frac{z_{ji}}{|z_{ji}|}\notag \\
&=&\sum_{2\le j\not=i}\frac{z_{ji}}{|z_{ji}|}\left[\left(-m_0+\frac{3}{2}k_j+\frac{nk_j^3+3k_1k_j^2-3k_jl_1}{2z_{1j}}\right)+\sum_{2\le k(\not=i,j)}\frac{(k_k-k_j)^3}{4z_{jk}}\frac{z_{ki}}{|z_{ki}|}\right]\notag \\
&=&\sum_{2\le j(\not=i)}\frac{z_{ji}}{|z_{ji}|}\left[\sum_{2\le k(\not=j)}\frac{(k_k-k_j)^3}{2|z_{kj}|}+\sum_{2\le k(\not=i,j)}\frac{(k_k-k_j)^3}{4z_{jk}}\frac{z_{ki}}{|z_{ki}|}\right],\label{eq:124}
\end{eqnarray}
where we we have used Eq.~(\ref{eq:c2=0}). After some lengthy, but straightforward computations, 
one can verify that (\ref{eq:124}) is further simplified to (\ref{eq:356}) up to the minus sign. 
This proves $\hat \omega_{(0)}=0$, as we desired to show.




\end{document}